# New Class of Dark Matter Objects and their Detection

C Sivaram and Kenath Arun

Indian Institute of Astrophysics, Bangalore

**Abstract:** About one-fourth of the universe is thought to consist of dark matter. Yet there is no clear understanding about the nature of these particles. Commonly discussed dark matter candidates includes the so called WIMPs or weakly interacting massive particles with masses from about $\sim 10 GeV$ to $1 TeV$. These particles can gravitate to form a new class of objects in dark matter halos or around the galactic centre. We study in some detail many properties of these objects (which are dark matter dominated and bound by their self gravity), their formation and possibilities of their detection. Implications of the presence of such objects for star formation are also discussed. These objects could provide the possibility of forming primordial black holes distinct from the usual Hawking black holes and they could also provide a scenario for short duration gamma ray bursts, avoiding the baryon load problem.

## I. Introduction

Current cosmological observations[1,2] imply that baryonic matter constitute only 4%, 25% is made up of so called dark matter (DM) considered to be massive particles (ranging from 10GeV to 1TeV) and about 70% of dark energy.[3] While dark energy is responsible for accelerating the universe with its negative pressure (like the cosmological constant), the presence of dark matter is so far been deduced mainly by flat rotation curve of galaxies and large velocity dispersion in galactic clusters. Other signatures of dark matter could be the release of radiation (gamma rays), when the particles annihilate. If they cluster and form gravitationally bound objects, these pairs of dark matter particles can annihilate through out these objects. These dark matter particle-antiparticle pairs can undergo annihilation and produce high energy gamma rays which could be detected. These high energy gamma rays could be a signature of this new class of objects.

Here we discuss the various dynamics of these objects made of pure dark matter particles[4], with the mass of dark matter particle varying form $10 GeV$ to $1 TeV$. As these dark matter particles, constituting the gravitationally bound object, have low non-thermal energies, we can consider the degeneracy pressure to be dominant.

The Chandrasekhar mass for these objects (which will be the upper limit on their mass) is given by:

$$M_{D(CH)} = \left(\frac{\hbar c}{G}\right)^{3/2} \frac{1}{m_D^2} \qquad \ldots (1)$$

Where, $m_D$ is the mass of the dark matter particle.

For a dark matter particle of mass $m_D \sim 100 GeV$, this works out to be:

$$M_D \approx 10^{27} g = 10^{-6} M_{sun} \qquad \ldots (2)$$

The size of these objects is given by (for the usual degenerate gas configuration):

$$M_D^{1/3} R = \frac{92\hbar^2}{G m_D^{8/3}} \qquad \ldots (3)$$

For the $10^{-6}$ solar mass object the size works out to be:

$$R \approx 10^5 cm \qquad \ldots (4)$$

And the corresponding velocity is:

$$v = \frac{\hbar n^{1/3}}{m_D} \qquad \ldots (5)$$

Where the number density $n \approx 10^{24} /cc$. Therefore we have $v \approx 10^3 cm/s$ $\qquad \ldots (6)$

The annihilation rate is $\sim n^2 \sigma v = 10^{15} /cc/s$ $\qquad \ldots (7)$

The cross section $\sigma \approx 10^{-36} cm^2$ (where we assume the standard wimp cross section)

Over the entire volume, the annihilation rate is given by: $\int_{R_C}^{R} n(r)^2 \sigma v. 4\pi r^2 dr$

The volume of the object $\sim (10^5)^3 cc$, from equation (4).

Therefore the total annihilation over the volume is $\sim 10^{30} /s$.

The annihilation of the dark matter particles takes place according to:[5,6] (The enhancement due to formation of Wimponium[7] is given in reference [3])

$$D + \overline{D} \rightarrow 2\gamma \qquad \ldots (8)$$

For dark matter particles of mass 100 GeV, each gamma photon has energy of $50 GeV \sim 0.1 ergs$. Hence the total energy radiated per second is

$$\dot{E} \sim (10^{30})(0.1) = 10^{29} ergs/s \qquad \ldots (9)$$

The total number of dark matter particles (of mass $m_D \sim 100 GeV$) present in the $10^{-6}$ solar mass object is:

$$\frac{M_D}{m_D} \approx 10^{49} \qquad \ldots (10)$$

If there are $\sim 10^{30}/s$ annihilations, then the life time of these DM objects is:

$$t_{life} \approx 10^{19} s \qquad \ldots (11)$$

As we see from the above result, the life span of these objects consisting predominantly of $m_D \sim 100 GeV$ dark matter particles is more than the age of the universe. Only about 10% of the mass of these objects would have evaporated in the Hubble time, thus such objects can still be detected by their gamma ray flux.

The flux from these objects at about 1kpc is given by:

$$f(1kpc) = \frac{10^{29}}{4\pi (3 \times 10^{21})^2} \approx 10^{-15} ergs/cm^2/s \qquad \ldots (12a)$$

These objects need not necessarily be confined to the galactic centre. They can be formed in the halos as well as over the galactic volume (in principle). The flux from an object one parsec away will be:

$$f(1pc) = \frac{10^{29}}{4\pi (3 \times 10^{18})^2} \approx 10^{-9} ergs/cm^2/s \qquad \ldots (12b)$$

This implies that on earth about 300 photons will be detected by a meter square detector over a year. Absence of such a flux enables constraints to be placed on the population abundance of such objects.[8]

A recent observation[9] of globular cluster M13 using MAGIC telescope sets an upper limit on the high energy gamma ray emission at $< 5.1 \times 10^{-12} cm^{-2} s^{-1}$. The flux from these objects at 1kpc (from equation (12a)) is of the order of $10^{-15} ergs cm^{-2} s^{-1}$.

The 100GeV dark matter particles annihilate to produce gamma photons of $\sim 0.1 ergs$. The flux of these gamma photons from M13 at 7.7kpc is given by:

$$f \sim 2 \times 10^{-16} cm^{-2} s^{-1} \qquad \ldots (13)$$

This sets a constraint on the number of these dark matter particles in the cluster M13 to less than about one in $10^6$.

As the dark matter particles annihilate over time, the mass and radius undergoes attrition over time. The rate at which the size changes with time:

$$\frac{dR}{dt} = \frac{dR}{dM}\frac{dM}{dt}$$

From equation (3) we can write,

$$\frac{dR}{dM} = \frac{M^{-4/3}}{3}\left(\frac{92\hbar^2}{Gm_D^{8/3}}\right) \qquad \ldots (14)$$

And the rate of loss of mass is given by:

$$\dot{M} = \frac{\dot{E}}{c^2} = \frac{10^{29} ergs/s}{c^2} \approx 10^8 g/s \qquad \ldots (15)$$

Where, from equation (9), $\dot{E} = 10^{29} ergs/s$, for $m_D \sim 100 GeV$ and:

$$\dot{R} = \frac{dR}{dM}\dot{M} = \frac{M^{-4/3}}{3}\left(\frac{92\hbar^2}{Gm_D^{8/3}}\right) \times \dot{M} \approx 10^{-19} cm/s \qquad \ldots (16)$$

The same analysis can be done for different dark matter particle mass. The results are tabulated below.

| $m_D$ (GeV) | $M_D(g)$ | $R(cm)$ | $t_{life}(s)$ | $\langle n^2 \sigma v \rangle V$ $(s^{-1})$ | $f(1kpc)$ $(ergs/cm^2/s)$ | $f(1pc)$ $(ergs/cm^2/s)$ | $\dot{M}$ $(g/s)$ | $\dot{R}$ $(cm/s)$ |
|---|---|---|---|---|---|---|---|---|
| 10 | $10^{29}$ | $6 \times 10^6$ | $10^{16}$ | $2 \times 10^{36}$ | $10^{-10}$ | $10^{-4}$ | $10^{13}$ | $10^{-13}$ |
| 100 | $10^{27}$ | $10^5$ | $10^{19}$ | $10^{30}$ | $10^{-15}$ | $10^{-9}$ | $10^8$ | $10^{-19}$ |
| 250 | $4 \times 10^{25}$ | $3 \times 10^3$ | $10^{21}$ | $10^{25}$ | $10^{-19}$ | $10^{-13}$ | $10^4$ | $10^{-23}$ |
| 500 | $10^{25}$ | $10^3$ | $5 \times 10^{22}$ | $10^{23}$ | $10^{-21}$ | $10^{-15}$ | $10^2$ | $10^{-25}$ |
| 1000 | $3 \times 10^{24}$ | $10^3$ | $3 \times 10^{23}$ | $10^{23}$ | $10^{-21}$ | $10^{-15}$ | $10^2$ | $10^{-25}$ |

From the table we can deduct the dependence of these parameters on the mass of dark matter particle. The rate of change of size of these objects is negligible and the luminosity becomes the Eddington luminosity only for dark matter particle mass of 10GeV and 100GeV.

The scaling of the relevant parameters with $m_D$ is as follows:

$$M_D \sim \frac{1}{m_D^2} \quad \ldots (17) \qquad\qquad R \sim \frac{1}{m_D^2} \quad \ldots (18)$$

$$\dot{M} \sim \frac{1}{m_D^6} \quad \ldots (19) \qquad\qquad \dot{R} \sim \frac{1}{m_D^6} \quad \ldots (20)$$

## 2. Galactic Halo radiation (from the above DM objects)

For these dark matter particles in the halo, assuming that they make up the dark matter, the kinetic energy of the dark matter particles associated with them is given by:

$$T = \frac{1}{2} m_D v^2 \qquad \ldots (21)$$

The dark matter particles form a collisionless gas held together by their self gravity.[6,10] If $M(r)$ is the mass contain within $r$, then

$$v^2 = \frac{2T}{m_D} = \frac{GM(r)}{r} \qquad \ldots (22)$$

From the continuity equation we have:

$$M(r) = \frac{2T}{Gm_D} = 4\pi r^2 \rho(r) dr \qquad \ldots (23)$$

The density within $r$ is then given by:

$$\rho(r) = \frac{T}{2\pi G m_D r^2} \qquad \ldots (24)$$

And the corresponding number density:

$$n(r) = \frac{T}{2\pi G m_D^2 r^2} \qquad \ldots (25)$$

The velocity in the halo is $\sim 3 \times 10^7 \, cm/s$ and for $100 GeV$ dark matter particle,

$$T \approx 5 \times 10^{-8} \, ergs \qquad \ldots (26)$$

Therefore the number density:

$$n(r) \approx 10^{-2} / cc \qquad \ldots (27)$$

Where $r \approx 30 kpc = 10^{23} cm$

And the corresponding density:

$$\rho(r) \approx 10^{-24} g / cc \qquad \ldots (28)$$

The annihilation per unit volume, $n(r)^2 \sigma v = 3 \times 10^{-34} / cc / s \qquad \ldots (29)$

And that over the entire volume $(V = 10^{69} cc)$ of the halo is given by:

$$\left(n(r)^2 \sigma v\right) V = \int_{R_C}^{R} n(r)^2 \sigma v . 4\pi r^2 dr = 3 \times 10^{35} / s \qquad \ldots (30)$$

The total energy released, as 50 GeV gamma photons, due to the total annihilation per second is:

$$\dot{E} \approx 10^{35} ergs / s \qquad \ldots (31)$$

The flux of photons received at earth is given by:

$$f = \frac{3 \times 10^{35}}{\left(10^{23}\right)^2} \approx 10^{-11} ergs / cm^2 / s \approx 10^{-9} / cm^2 / s \qquad \ldots (32)$$

This would translate to about $10^3$ / year photons detected for a 10 square meter detector!

### 3. Radiation from Galactic Clusters

In the case of galaxy clusters, the velocity dispersion is $\sim 2 \times 10^8 cm / s$ and thus for a $100 GeV$ dark matter particle,

$$T \approx 2 \times 10^{-6} ergs \qquad \ldots (33)$$

Therefore the number density:

$$n(r) \approx 5 \times 10^{-5} / cc \qquad \ldots (34)$$

Where $r \approx 1 Mpc = 3 \times 10^{24} cm$ is the typical galactic cluster size.

And the corresponding density:

$$\rho(r) \approx 5 \times 10^{-27} g / cc \qquad \ldots (35)$$

The annihilation per unit volume, $n^2 \sigma v = 5 \times 10^{-38} / cc / s$ ... (36)

And that over the entire volume $(V = 10^{74} cc)$ of the cluster is:

$$(n^2 \sigma v)V = 5 \times 10^{36} / s \qquad \ldots (37)$$

The total energy released, as 50 GeV gamma photons, due to the total annihilation per second is: $\dot{E} \approx 10^{36} ergs/s$ ... (38)

$$\text{The flux of photons at earth} = \frac{5 \times 10^{36}}{(10^{26})^2} \approx 10^{-17} / cm^2 / s \qquad \ldots (39)$$

(At a distance $D = 60 Mpc \sim 10^{26} cm$)

For a $100 m^2$ detector this works out to be about $10^{-4} / year$.

## 4. Limiting luminosity from these DM Dominated Objects

The Eddington luminosity associated with a mass $M$ is given by:

$$L_{Edd} = \frac{4\pi GMc}{\sigma_T} \qquad \ldots (40)$$

Where $\sigma_T$ is the Thomson cross section.

For 100 GeV dark matter particle, from equation (2), $M_D \approx 10^{27} g$. The corresponding Eddington luminosity is:

$$L_{Edd} \approx 4 \times 10^{30} ergs/s \qquad \ldots (41)$$

From the table, we can see that the luminosity for the object is almost equal to the Eddington luminosity. This holds true (luminosity of the object equal to the Eddington luminosity) only for dark matter particle mass of 10GeV and 100GeV.

## 5. Clustering of these Objects at Galactic Centre

The galactic centre contains a dense region with mass of $\sim 3 \times 10^6$ solar mass. If this region is composed of these dark matter objects with the dark matter particle mass of 100 GeV, then there should be $\sim 10^{12}$ such objects.

This would then imply a total energy radiated per second:

$$\dot{E} \sim (10^{29})(10^{12}) = 10^{41} ergs/s \qquad \ldots (42)$$

(From equation (9), the energy given out per unit time for each of such objects is $\sim 10^{29} \, ergs/s$.)

The energy radiated per second from such an object $\left(10^{41} \, ergs/s\right)$ hasn't been detected from the centre of the galaxy, hence such a scenario can be ruled out.

## 6. Rotation of Dark Matter Objects and possible Gravitational Radiation form them

These dark matter objects can rotate about their axis and the break up speed is given by:

$$\frac{v^2}{R} = \frac{GM}{R^2} \qquad \ldots (43)$$

Where $M = 10^{27} g$ and $R = 10^5 \, cm$ for dark matter particles of mass $m_D = 100 \, GeV$. The break up speed for such objects is:

$$v \approx 3 \times 10^7 \, cm/s \qquad \ldots (43a)$$

The corresponding frequency of rotation and the period is given by:

$$\omega = \sqrt{\frac{GM}{R^3}} \approx 100 \, Hz$$
$$P = \frac{2\pi}{\omega} \approx 0.01 s \qquad \ldots (44)$$

This gives a new class of objects with high frequency like pulsars. But unlike pulsars they do not release magnetic dipole radiation but lose energy through emission of gravitational waves for objects with high ellipticity $(e)$.

The gravitational wave radiation is given by: $\dot{E}_G = \frac{32G}{5c^5} I_M^2 \omega^6 e^2 \qquad \ldots (45)$

For objects with low ellipticity the rate of emission of gravitational radiation is less and hence their life time $\left(t_{life} = \frac{B.E}{\dot{E}_G}\right)$ is much greater then the Hubble age.

## 7. Can AGN Cores be made up of these DM Objects?

For AGNs ($10^{12}$ solar mass) to be made up of such objects, there has to be $10^{18}$ such objects. The total energy radiated per second would be:

$$\dot{E} \sim \left(10^{29}\right)\left(10^{18}\right) = 10^{47} \, ergs/s \qquad \ldots (46)$$

The flux due to this on earth (at $10^{27}$ cm) will be:

$$f = \frac{10^{47}}{4\pi(10^{27})^2} \approx 10^{-8} \, ergs/cm^2/s \qquad \ldots (47)$$

## 8. Possible scenario for formation of Black Holes by the merger of these objects

The time taken for these objects to merge is given by:

$$t_{merge} = \frac{v^3}{NG^2M^2 \ln \Lambda} \qquad \ldots (48)$$

Where, $N$ is the number density of the objects, $\ln \Lambda \sim 10$ and the velocity $v \sim 10^8 \, cm/s$

For these objects to merge in Hubble time, the number density of these objects at the galactic centre should be:

$$N = \frac{v^3}{t_{merge} G^2 M^2 \ln \Lambda} \approx 10^{21}/(pc)^3 \qquad \ldots (49)$$

Since such large densities are untenable, such scenario for black hole formation at galactic centres is ruled out.

## 9. Possible scenario for star formation with such Objects

The above discussion applies to the final state of these dark matter objects. Now we look at these objects while they are contracting from the primordial mixture of dark matter and baryonic matter. Early galaxies have been detected at a redshift of about 10. This[11,12] indicates that the collapse would have happened earlier, say at a $z \approx 20$.

The density at that redshift is given by:

$$\rho = 0.25 \rho_0 (1+z)^3 \qquad \ldots (50)$$

Where, 0.25 is the fraction of dark matter of the total energy density and $\rho_0 \approx 10 keV/cc$ is the present energy density.

For dark matter particles of mass $m_D = 100 GeV$, the number density of dark matter particles at $z \approx 20$ is given by:

$$n = 0.25 \left(\frac{10 keV}{100 GeV}\right)(1+z)^3 \approx 10^{-3}/cc \qquad \ldots (51)$$

And if the density of the initial cloud which collapsed is about 10 times the ambient density then $n = 10^{-2}/cc$. The mass density of the dark matter particles of mass $m_D = 100 GeV$ is:

$$\rho = (10^{-22} g)(10^{-2}/cc) = 10^{-24} g/cc \qquad \text{... (52)}$$

The time taken for the initial cloud (of dark matter and baryonic matter) to collapse is given by:

$$t_{collapse} \sim \frac{1}{\sqrt{G\rho}} = 10^{15} s \qquad \text{... (53)}$$

As the cloud collapses, the dark matter particles decouple with radiation where as the baryonic matter will be heated up. The temperature of the baryonic matter is given by:

$$T = \frac{GM_D m_p}{k_B R} \qquad \text{... (54)}$$

Where, the mass and the size of the object for 100GeV dark matter particles is given by equations (2) and (4) as $M_D \approx 10^{27} g$ and $R \approx 10^5 cm$. And $m_p$ is the proton mass.

The temperature then works out to be:

$$T \approx 10^7 K \qquad \text{... (55)}$$

At this temperature, thermonuclear reactions will start.

This scenario of baryonic matter collapsing along with the dark matter particles could thus lead to the formation of stars.

There will be gamma ray emissions as the thermonuclear reaction proceeds in the star. But these will be MeV gamma rays as opposed to the 100GeV gamma rays coming from the dark matter objects.

The energy density due to radiation will be comparable to the collapsing gas's energy density. That is:

$$aT^4 = nk_B T \qquad \text{... (56)}$$

For $T \approx 10^7 K$, the number density will be

$$n \approx 10^{22}/cc \qquad \text{... (57)}$$

This is the number density at the stellar cores.

As for the possible masses of these objects, we can estimate as follows:

The energy released during the collapse is given by:

$$E_{grav} = \frac{GM_D^2}{R} \approx 10^{42} \, ergs \qquad \ldots (58)$$

This released gravitational potential energy will heat up the gas accreting on to the objects (DM would be thermally decoupled)

$$MR_g T = 10^{42} \, ergs \qquad \ldots (59)$$

Where $R_g$ is the universal gas constant and from equation (56), temperature $T \approx 10^7 \, K$.

The mass of these objects is therefore, $M = 10^{26} \, g$, which is sub-stellar mass.

The volume of the object, $V = \dfrac{M}{nm_D} \approx 10^{27} \, cc$

The annihilation rate, $n^2 \sigma V \approx 10^{31} \, ergs/s$, which is about $10^{-2} - 10^{-3}$ solar luminosity.

The expected temperature of these objects will be about $10^4 \, K$, so they will be blue stars.

The life time of such objects will be:

$$t_{life} = \frac{10^{26} \, g}{10^{12} \, g/s} \sim 10^{14} \, s \qquad \ldots (60)$$

This could give rise to a new class of stellar objects, but is based on assumptions that only one dark matter object is involved and no mergers. Could be very rare and difficult to detect unless within a few parsecs.

## 10. Accretion of Dark Matter Particles into Black Hole

The dark matter particles of mass $m_D$ can be accreted[13] into a black hole of mass $M_{BH} \approx 3 \times 10^6 \, M_{sun}$ as in the centre of our galaxy. The rate of accretion by the black hole is given by:

$$\dot{M} = 4\pi (fR_S)^2 nm_D \left( \frac{GM_{BH}}{fR_S} \right) \qquad \ldots (61)$$

Where, $n \approx 10^{-3} - 10^{-4}/cc$ is the ambient number density, $R_S \approx 10^{11} cm$ is the Schwarzschild radius for the million solar mass black hole and $f \sim 100$ is the extent of the ambient gas from the black hole.

For a 100 GeV dark matter particle, the accretion rate is:

$$\dot{M} \approx 10^{17} g/s \qquad \ldots (62)$$

If all the accreted particles undergo annihilation, then the energy released is given by:

$$\dot{E} = \dot{M}c^2 \approx 10^{38} ergs/s \qquad \ldots (63)$$

The corresponding flux of photons, each of energy 50GeV (~0.1ergs) at 10kpc (galactic centre), is given by:

$$f = \frac{\dot{n}c^2}{4\pi R^2} = \frac{10^{37}}{4\pi(3\times 10^{22})^2} \approx 10^{-9} ergs/cm/s \qquad \ldots (64)$$

For a $10m^2$ detector, this corresponds to about $3\times 10^7$ photons/year or ~ one photon/second.

And at 1Mpc, the flux $f = \frac{10^{37}}{4\pi(3\times 10^{24})^2} \approx 10^{-13} ergs/cm/s \qquad \ldots (65)$

And for a $10m^2$ detector, this corresponds to ~ 300 ergs/year or ~ 3000 photon/year.

## 11. Collapse of these DM Objects to form Micro BHs: Possible Alternate Scenario for short Duration GRBs

The Chandrasekhar mass (upper limit) for these DM objects is given by equation (1) as:

$$M_{D(CH)} = \left(\frac{\hbar c}{G}\right)^{3/2} \frac{1}{m_D^2}$$

If their mass exceeds this limit, they will collapse to form black holes of size given by:

$$R_S = \frac{2GM}{c^2} \approx 1 cm \qquad \ldots (66)$$

(For DM particles of mass $m_D = 100 GeV$)

The energy released during the collapse is given by:

$$E = \frac{GM^2}{R} \approx 10^{48} ergs \qquad \ldots (67)$$

This energy is released in the form of gravitational waves.

If equal amount of baryonic matter collapses along with the dark matter to form the black hole, then the baryonic matter will be heated up according to:

$$MR_g T = 10^{48} \Rightarrow T \approx 10^{12} K \quad \ldots (68)$$

This corresponds to gamma ray frequency. Since the mass heated up is $\sim 10^{-6}$ solar mass, in this scenario, the 'Baryon Load' problem seems ameliorated.

The time scale of the gamma ray burst is given by:

$$t_{burst} = \sqrt{\frac{R^3}{GM}} \approx 0.01s \quad \ldots (69)$$

This could be an alternative scenario for short duration $(0.1 - 0.01s)$ sub-luminous gamma ray bursts. This is another way in which primordial black holes ($<$ stellar mass) can form, apart from Hawking black holes.

## 12. Binary System of these DM Objects

These dark matter objects can form binary systems, with each of mass $10^{27} g$ and size of $10^5 cm$. If their separation is about tem times their size, then the period is given by:

$$GMP^2 = 4\pi^2 R^3$$
$$P \approx 1s \quad \ldots (70)$$

If these binary systems are present at a distance of 20Mpc, the corresponding strain at earth due to gravitational radiation emission from them is:

$$h = \frac{2GE}{rc^4} \quad \ldots (71)$$

Where $E \sim 10^{54} ergs$ is the energy released. The strain then becomes:

$$h \approx 3 \times 10^{-21} \quad \ldots (72)$$

The flux of the gravitational waves is given by:

$$F_{GW} = \frac{c^3}{32\pi G}\left(\frac{dh}{dt}\right)^2 \approx 4 \times 10^{-6} W/m^2 \quad \ldots (73)$$

The strain due to gravitational radiation emission from these binaries at different distances from earth and the corresponding flux is given as follows:

$h \approx 6 \times 10^{-18}$; $F_{GW} \approx 10 \, W/m^2$ at 10kpc

$h \approx 6 \times 10^{-17}$; $F_{GW} \approx 10^3 \, W/m^2$ at 1kpc

$h \approx 6 \times 10^{-15}$; $F_{GW} \approx 10^9 \, W/m^2$ at 1pc

**13. Integrated Gamma Ray Flux**

The dark matter particles decoupled from radiation at early redshift and collapsed to form these dark matter objects, hence contributing to the background gamma ray flux. The total fluence from present $(z=0)$ to $z$ is given by:

$$F_I = \frac{N t_d c n_0}{H_0} \int_0^z \frac{dz}{(1+z)^2 (1+\Omega z)^{1/2}} \qquad \ldots (74)$$

Where, N is the gamma rays emitted by each object per second

$t_d$ is the average life time of these objects

$H_0 = 70 \, km/s/Mpc$ is the Hubble constant at $z=0$

$n_0$ is the number density of these objects

If half of the dark matter in the universe collapses to form these objects, then

$$n_0 = \frac{1}{2} \frac{0.26 M_{total}}{M_D (2\pi^2 R_H^3)} \qquad \ldots (75)$$

For a flat universe $(\Omega = 1)$, the total energy density of the universe:

$$\frac{M_{total}}{(2\pi^2 R_H^3)} \approx 10^{-29} \, g/cc \qquad \ldots (76)$$

The average life time of these objects form equation (60) is, $t_d \approx 10^{14} \, s$

For a 10GeV dark matter particle the mass of the object $M_D \approx 10^{29} \, g$ (from the table). Hence the background number density of these objects is constraint to be:

$$n_0 < 10^{-4} \, pc^{-3} \qquad \ldots (77)$$

And the gamma rays emitted by each object per second (from the table): $N = 10^{36} / s$

The corresponding fluence is given by:

$$F_I = 10^{13} cm^{-2} \left(\frac{N(s^{-1})}{10^{36}}\right)\left(\frac{t_d(s)}{10^{14}}\right)\left(\frac{c(cm/s)}{3 \times 10^{10}}\right)\left(\frac{n_0(cm^{-3})}{10^{-59}}\right)\left(\frac{70(km/s/Mpc)}{H_0}\right)\left(\frac{2}{3}\right) \quad \ldots (78)$$

The flux is therefore given by:

$$f_I = \frac{F_I}{t_H} \approx 10 cm^{-2} s^{-1} \quad \ldots (79)$$

Where $t_H \approx 10^{10}$ years is the Hubble time.

For 100GeV DM particle, $M_D \approx 10^{27} g$; $n_0 \approx 10^{-57} /cc$; $N = 10^{30} /s$

And the corresponding flux: $f_I = 10^{-3} cm^{-2} s^{-1}$ ... (80)

For 250GeV DM particle, $M_D \approx 10^{25} g$; $n_0 \approx 10^{-55} /cc$; $N = 10^{25} /s$

And the corresponding flux: $f_I = 10^{-6} cm^{-2} s^{-1}$ ... (81)

This would put a limit on the background density of these objects.

Future observations could constrain the masses of the dark matter particles predicted by various particle physics models.[14]

## 14. Concluding Remarks

In this paper we look at a new class of objects formed by the collapse of dark matter particles and elucidated their various physical properties and observable aspects associated with them. We have analysed the effects of the mass of individual dark matter particles on the mass, annihilation rate, mass loss, etc of these dark matter objects. In a certain range of parameters their life times are comparable to the Hubble age and they could be observable as gamma ray point sources (like Hawking black holes, but with distinctly different signatures). We have also explored the possibilities of star formation and black hole formation from accretion onto and mergers of these dark matter objects. These objects could provide the possibility of forming primordial black holes (sub-stellar) distinct from the usual Hawking black holes. As shown in section 11, they could also provide a scenario for short duration gamma ray bursts, avoiding the baryon load problem. The possible integrated back ground flux from these objects to the cosmological gamma ray flux is also evaluated.

**Reference:**

1. Perlmutter S et al, Nature 391, 5, 1998
2. Krauss L M, Ap. J., 604, 91, 2004
3. Zioutask S et al, Science, 306, 2004
4. Diemand J et al, Nature, 433, 389, 2005
5. Guzzo L et al, Modern Cosmology, eds. S Bonometto et al, Bristol, UK, 344
6. Spergel D N et al, astro-ph/0603449, 2006
7. March-Russell J and West S M, Phys. Lett. B, 676, 133, 2009
8. White S, SDM, Cosmology and Large Scale Structure, 349, 1996
9. Anderhub H et al, Ap. J., 702, 266, 2009
10. Navarro J and White S, MNRAS, 265, 271, 1993; Navarro J et al, PRD, 69, 2004
11. Van der Bosch F C, MNRAS, 327, 1334, 2001
12. Firmani C and Avila-Reese V, astro-ph/0303543, 2003
13. Seljak V et al, PRD, D71, 103515, 2005
14. Cembranus J et al, PRL, 99, 191301, 2007